\documentclass[12pt,nofootinbib]{revtex4}
\usepackage{amsmath}
\usepackage{bm}
\usepackage{bbm}
\usepackage{amssymb}
\usepackage{amsthm}
\usepackage{amsfonts}
\usepackage[latin1]{inputenc}
\usepackage{graphicx}
\usepackage{epstopdf}

\def\TT{\mathtt{T}}

\newtheorem{cla}{Claim}

\newtheorem*{thm*}{Theorem}
\newtheorem{defi}{Definition}

\numberwithin{equation}{section}

\def\RRR{{\mathbb R}}

\def\NNN{{\mathbb N}}

\def\GG{{\cal G}}

\def\0{\emptyset}
\def\z{\zeta}
\def\L{\Lambda}

\def\d{\delta}

\def\g{\gamma}

\def\e{\varepsilon}

\def\G{\Gamma}
\def\p{\pi}

\def\o{\omega}

\def\bz{{\bm z}}

\def\bv{{\bm v}}

\def\bt{{\bm t}}

\def\bze{{\bm\zeta}}

\def\bet{{\bm \eta}}

\newcommand{\pa}{\partial}

\def\be{\begin{equation}}
\def\ee{\end{equation}}
\def\bea{\begin{eqnarray}}
\def\eea{\end{eqnarray}}

\def\nn{\nonumber}

\begin{document}

\title{Kinetic Theory of Cluster Dynamics}

\maketitle

\begin{center}
Robert I. A. Patterson

{\em Weierstrass Institute Berlin \\ 
Mohrenstr. 39, 10117 Berlin, Germany \\ 
}
\texttt{\small robert.patterson@wias-berlin.de}

\bigskip

Sergio Simonella

{\em Zentrum Mathematik, TU M\"{u}nchen \\ 
Boltzmannstr. 3, 85748 Garching, Germany \\ 
}
\texttt{\small s.simonella@tum.de}

\bigskip

Wolfgang Wagner

{\em Weierstrass Institute Berlin \\ 
Mohrenstr. 39, 10117 Berlin, Germany \\ 
}
\texttt{\small wolfgang.wagner@wias-berlin.de}

\end{center}

\bigskip

\noindent {\small In a Newtonian system with localized interactions the whole set of particles is naturally decomposed 
into dynamical clusters, defined as finite groups of particles having an influence on each other's trajectory during 
a given interval of time. For an ideal gas with short--range intermolecular force, we provide a description of
the cluster size distribution in terms of the reduced Boltzmann density. In the simplified context of 
Maxwell molecules, we show that a macroscopic fraction of the gas forms a giant component in finite kinetic time. 
The critical index of this phase transition is in agreement with previous numerical
results on the elastic billiard.
}

\bigskip

\noindent {\small {\em Keywords:} low--density gas; Boltzmann equation; cluster dynamics; Maxwell molecules.}


\section{Introduction} \label{sec:int}

As a proposal to gain insight on the statistical properties of large systems in a gaseous phase,
N. Bogolyubov suggested to investigate a simple notion of {\em cluster decomposition} characterizing the collisional 
dynamics \cite{Bo46}. When the evolution is determined by a sequence of single, distinct, two--body interactions, 
a natural partition of the system can be defined in terms of {\em groups} of particles connected by a chain of collisions,
so that a ``cluster'' consists of elements having affected each other's trajectory.

This notion has been developed later on, in connection with the problem of the Hamiltonian
dynamics of an infinite system. In a Newtonian system, particles with rapidly increasing energies at infinity may generate
instantaneous collapses for special initial configurations \cite{La68}. Mathematically, one needs to prove that such
initial data form a set of measure zero in the phase space of infinitely many particles. In fact, one possible strategy to 
construct the dynamics is to show that, at properly fixed time, the system splits up into an infinite number of
clusters which are moving {\em independently} as finite-dimensional dynamical systems. After some random interval
of time, the partition into independent clusters changes, and one iterates the procedure. This dynamics is known as
{\em cluster dynamics} and its existence has been proved first in \cite{Si72} for some one--dimensional models
(see \cite{Si74} for generalizations).

In more recent years, the statistical properties of cluster dynamics of a system obeying Newton's law
have been studied numerically \cite{GKBSZ08}. In this reference, the authors focus on the frictionless elastic 
billiard in a square two--dimensional box with reflecting walls,
and show that the dynamics undergoes a phase transition. This occurs
in a way reminiscent of problems in percolation theory. Namely, the maximal (largest) cluster starts to 
increase dramatically at some critical time. At the critical time, the fraction of mass in the maximal cluster is 
rather small ($\sim 7\%$ for $5000$ disks at small volume density). After the critical time, it approaches 
the total mass of the system with exponential rate. 
Moreover, the transition is distinguished by a 
power--law behaviour for the cluster size distribution with exponent $5/2$. Such critical index is believed to be 
{\em universal}, since it has been observed for several different models (see also \cite{McFB10}).

The cluster dynamics concept, together with the above described statistical behaviour, appear as well in a 
number of applied papers, e.g. geophysics, economics, plasma physics: see \cite{GKBSZ08} and references therein.

Kinetic theory often provides successful methods for the computation of microscopic quantities related to 
properties of the dynamical system, for instance Lyapunov exponents or Kolmogorov--Sinai entropies 
\cite{vZvBD00, vBDDP97, Co97, Sp91}.
In the present work, we are concerned with the cluster dynamics of an ideal gas where the kinetic description 
of Boltzmann based on molecular chaos applies. 

Our setting is given by a density function $f = f(x,v,t)$ describing 
the amount of molecules having position $x \in\L \subset \RRR^d$ and velocity $v\in \RRR^d$ at time $t$, and evolution 
ruled by
\be
(\pa_t+v\cdot \nabla_x)f
= \int_{\RRR^d\times S^{d-1}} dv_1 d\o \ B(v-v_1,\o) \Big\{f'\,f'_1-
f\,f_1\Big\}\;,
\label{BE}
\ee
where $f = f(x,v,t), f_1 = f(x,v_1,t), f'= f(x,v',t), f'_1= f(x,v_1',t)$, $(v,v_1)$ is a pair of 
velocities in incoming collision configuration  and $(v',v_1')$ is the corresponding pair of outgoing 
velocities when the scattering vector is $\o$:
\be
\begin{cases}
\displaystyle v'=v-\o [\o\cdot(v-v_1)] \\
\displaystyle  v_1'=v_1+\o[\o\cdot(v-v_1)]
\end{cases}\label{eq:coll}\;.
\ee
The time--zero density $f(x,v,0) = f_0(x,v)$ is fixed. For simplicity,
the gas moves in the square $d-$dimensional box $\L$ of volume $1$, with reflecting boundary conditions.
The microscopic potential is assumed to be short--ranged and the cross--section $B$
satisfies $\int d\o \,B(v-v_1,\o) = a(|v-v_1|) < \infty$ (``Grad's cut--off assumption''). 

The precise connection with a dynamical system of $N$ particles interacting at mutual distance $\e$, such as the
one studied in \cite{GKBSZ08}, can be established locally in the low--density limit 
\be
N \to \infty\;, \ \ N\e^{d-1} \simeq~1\;,
\label{eq:BG}
\ee
(``Boltzmann--Grad regime'') as the  convergence of correlation functions to the solution of the Boltzmann
equation \cite{La75} (see also \cite{Uc88, Sp91, CIP94, GSRT12, PSS13}). 
In the regime \eqref{eq:BG}, the gas is so dilute that only two--body collisions 
are relevant. Furthermore, the collisions are completely localized in space and time. The limit transition \eqref{eq:BG}
explains the microscopic origin of irreversible behaviour \cite{Gr58}.

Our purpose here is to describe how the cluster size distribution is constructed from the solution to the Boltzmann
equation. This is done in Section \ref{sec:dc} by means of a suitable tree graph expansion, which is
inspired by previously known formulas representing the Boltzmann density as a sum over collision sequences \cite{Wi51}.
In Section \ref{sec:BG}, we indicate how to derive formally the introduced expressions
as the limiting cluster distributions of a system of hard spheres in the Boltzmann--Grad scaling.
Finally, in Section \ref{sec:Max}, we restrict to the simplest nontrivial (and paradigmatic) case in kinetic theory,
i.e. the model of Maxwellian molecules. We show that the cluster distribution exhibits a phase transition characterized by a breakdown of the normalization condition at finite time.
This implies that the ``percolation'' {\em survives} in the Boltzmann--Grad limit,
with same qualitative behaviour and same critical index of the elastic billiard analyzed in \cite{GKBSZ08}.

\section{cluster distributions} \label{sec:dc}

\subsection{Bogolyubov clusters} \label{sec:dc1}

We start with a formal notion of cluster. Let $t$ be a given positive time.
\begin{defi}\ \label{def:BClu}

\noindent (i) Two particles are $t-$neighbours if they collided during the time interval $[0,t]$.

\noindent (ii) A Bogolyubov $t-$cluster is any connected component of the neighbour relation (i).
\end{defi}

\noindent The definition can be generalized to generic time intervals $[s,s+t]$. However, in what follows we will
study the notion of $t-$cluster only, which is no restriction, and drop often the $t$-dependence
in the nomenclature.

Notice that each particle has collided with {\em at least} one other particle of its Bogolyubov cluster,
while it has {\em never} collided with particles outside the cluster, within the time interval $[0,t]$. In particular,
if $t=0$, {\em any} particle of the gas forms a singleton (cluster of size $1$). At $t>0$, the mass of singletons
starts to decrease and clusters of size $k=2,3,\cdots$ start to appear. We therefore expect to see (and do
observe in the experiments) some ``smooth'' exponential distribution in the cluster size.

\subsection{Backward clusters} \label{sec:dc2}

In Reference \cite{APST14}, the solution of \eqref{BE} has been expanded in terms of a sum of type
\be
f=\sum_{n=0}^\infty \sum_{\Gamma_n} \,f^{\G_n}\;,
\label{exp}
\ee
where $\G_0 = \emptyset$, $\G_n = (k_1,k_2,\cdots,k_n)$ and $k_1\in \{1\} , k_2\in \{1,2\}, \cdots, k_n \in \{1,2, \cdots, n\}$.
The sequences $\G_n$ are in one--to--one correspondence with binary tree graphs, e.g.
\be
\includegraphics[width=4in]{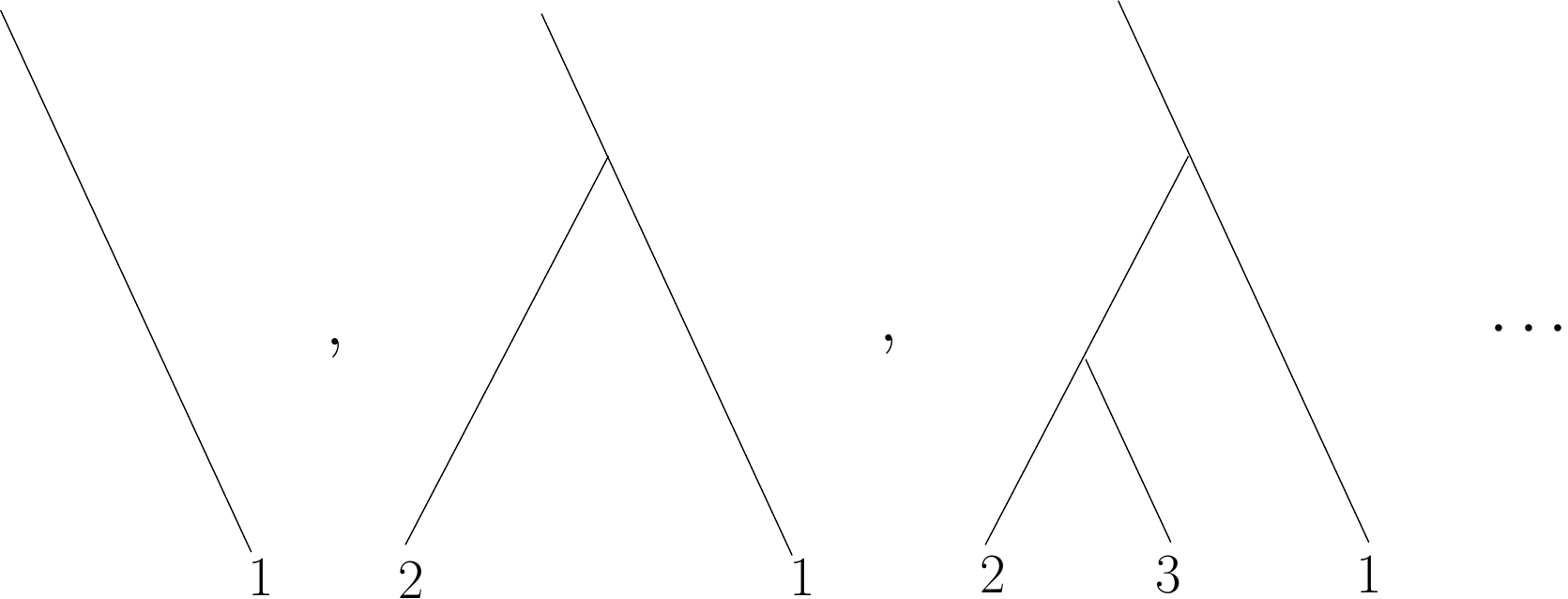} \label{eq:figI2}
\ee
for $n=1, 2, 3 \cdots$ respectively. In \eqref{exp}, $f^{\G_n}$ is interpreted as the contribution to the 
probability density $f$ due to the event: {\it the backward cluster of $1$ has structure $\G_n$}. 
By ``backward cluster'' we mean here the group of particles involved directly or indirectly in the 
backwards--in--time dynamics of particle $1$. Operatively, in a numerical experiment, we select particle
$1$ at time $t$, run the system backwards in time, and collect all the particles which collide
with $1$ and with ``descendants'' of $1$ in the backwards dynamics, following \eqref{eq:figI2}.
In other words, \eqref{exp} is an expansion on sequences of real collisions\footnote{Observe that recollisions
(e.g. the pair $(1,2)$ colliding twice in the backward history), certainly possible in an experiment,
do not affect the notion of backward cluster,
which is based on sequences of collisions involving at least one ``new'' particle; see \cite{APST14} for details
on the numerical procedure.}.

Formulas of this kind have been previously studied in the context of Maxwellian molecules with cut--off under the 
name of Wild sums \cite{Wi51, Mc66, CCG00}, and are written in \cite{APST14} for a gas of hard spheres in
a homogeneous state. It is not difficult to generalize such a representation to inhomogeneous states and 
general interactions. The formula for $f^{\G_n}$ reads
\bea
\label{eq:fjexp2}
&&f^{\G_n}(x_1,v_1,t)=  \int_0^t dt_1\int_0^{t_1} dt_2  \cdots \int_0^{t_{n-1}}dt_n \int_{\RRR^{nd}} 
dv_{2} \cdots dv_{1+n} e^{-\int_{t_{1}}^{t} ds\, R_{1}(\z_1(s),s)}\,
 \nn \\ 
&& \ \ \ \left(\prod_{r=1}^n  
\int_{S^{d-1}} d\o_r \, B(\eta^{r-1}_{k_r} -v_{1+r}\,,\,\o_r) \,
e^{-\int_{t_{r+1}}^{t_{r}} ds\, R_{1+r}(\bze_{r+1}(s),s)}\right)
f_{0}^{\otimes(1+n)} (\bze_{n+1}(0))\;, \nn\\
\eea
where: 

-- $\bze_k = (\z_1,\cdots,\z_k)$, $\z_i = (\xi_i,\eta_i) = $ (position, velocity);

-- $f_0$ is the initial density (and $f_{0}^{\otimes(1+n)} (\bze_{n+1})=f_0(\z_1)\cdots f_0(\z_{n+1})$);

-- the ``free--flight rate'' of $k$ particles $R_k$ is given by
\be
R_k (\bze_k,s) = \sum_{i=1}^k R(\z_i,s) \label{eq:Rjdef}
\ee
where the function $R$ depends on the solution $f(s)$ of the Boltzmann equation itself:
\be
R (x,v,t)=  \int_{\RRR^{d}\times S^{d-1}} dv_* d\o \ B(v-v_*,\o) f(x,v_*,t)\;;
\label{nu}
\ee

-- the ``{\em trajectory} of the backward cluster'' $s \to \bze_{r+1}(s)$ is constructed as follows:

(a) fix $x_1,v_1,t,\G_n, t_1,v_2,\o_1, \cdots, t_{n},v_{n+1},\o_{n}$, with $t  > t_1 > \cdots > t_n >  0\;;$

(b) construct first the sequence of velocities $\bet^r$, $r=0, \cdots, n$, defined iteratively by:
$$
\bet^0=\bv_{n+1} = (v_1,\cdots,v_{n+1})\;,    \quad \bet^r=( \eta_1^{r-1},\cdots, \eta_{k_r}', \cdots,\eta_{r+1}',\cdots,\eta_{n+1}^{r-1})   \quad r\geq 1
$$
where, at step $r$,  the pair  $\eta_{k_r}'  , \eta_{r+1}'$ are the pre--collisional velocities 
(in the collision with impact vector $\o_r$) of the pair $\eta_{k_r}^{r-1},  \eta_{r+1}^{r-1} = v_{r+1}$ 
(which are post--collisional, as ensured by the fact that $B(v-v_1,\o)\neq 0$ only for $(v-v_1)\cdot\o >0$),
according to the transformation \eqref{eq:coll};

(c) construct the trajectory of the backward cluster iteratively by 
$$
\xi_1(s) = x_1 -v_1(t-s)\;, \quad \eta_1(s) = v_1 \quad\quad\quad\quad\quad\quad s\in (t_1,t)\;,    $$ and, for $r\geq 1$ and $i=1,\cdots,r+1$,
$$ \xi_i(s) = \xi_i(t_r)-\eta_i^r(t_r-s)\;, \quad \eta_i(s) = \eta_i^r \quad\quad\quad\quad\quad\quad s\in (t_{r+1},t_r)\;,
$$
with the convention $t_{n+1} = 0$.

The term $n=0$ in \eqref{eq:fjexp2} is
\be
f^{\emptyset}(x,v,t) = e^{-\int_0^t ds R (x-v(t-s),v,s) } f_0(x-vt,v)
\label{eq:ffr}
\ee
and $\int dx dv  \,f^{\emptyset}(x,v,t) = \ $ density of free particles in $(0,t)$.
Similarly, $\int dx dv \sum_{\Gamma_n} \,f^{\G_n}(x,v,t) = \ $density of particles having a backward cluster 
of size $n$ in $(0,t)$.


\subsection{Symmetrization} \label{sec:dc3}

We can see $f^{\G_n}$ as an integral over trajectories of a Markov process with $n$ collisions in specified order.
Each trajectory has probability density given by the integrand function, that is:
\bea
&& f_{0}^{\otimes(1+n)} = \mbox{initial distribution of $n+1$ particles}\;;\nn\\
&& B(\eta^{r-1}_{k_r} -v_{1+r}\,,\,\o_r) = \mbox{transition kernel of the collision $(\eta_{k_r}^{r-1},  v_{r+1})\to
(\eta_{k_r}'  , \eta_{r+1}')$}\;;\nn\\
&& e^{-\int_{t_{r+1}}^{t_{r}} ds\, R_{1+r}(\bze_{r+1}(s),s)} = \mbox{probability of free flight of $r+1$ particles
in $(t_{r+1},t_{r})$, } \nn\\
&&\ \ \ \ \ \ \ \ \ \ \ \ \ \ \ \ \ \ \ \ \ \ \ \ \ \ \ \ \ \ \ \mbox{conditioned to the configuration $\bze_{r+1}(t_r)$}\;.\nn\\
\label{eq:pdtr}
\eea
We remind that, in a backward cluster, the trajectory of particle $i$ is specified {\em only} in the time interval $(0,t_{i-1})$, where 
$t  > t_1 > \cdots > t_n >  0$.

The density of Bogolyubov clusters can be obtained by ``adding'' the missing information, namely the {\em future history}
of the particles $2,3,\cdots$ in the time intervals $(t_1,t), (t_2,t),\cdots$ respectively, together with the complete history 
of the particles with whom they collide.
This amounts, for instance, to extend in the future,
by free motion, the trajectories of $2,3,\cdots$
\be
\includegraphics[width=4in]{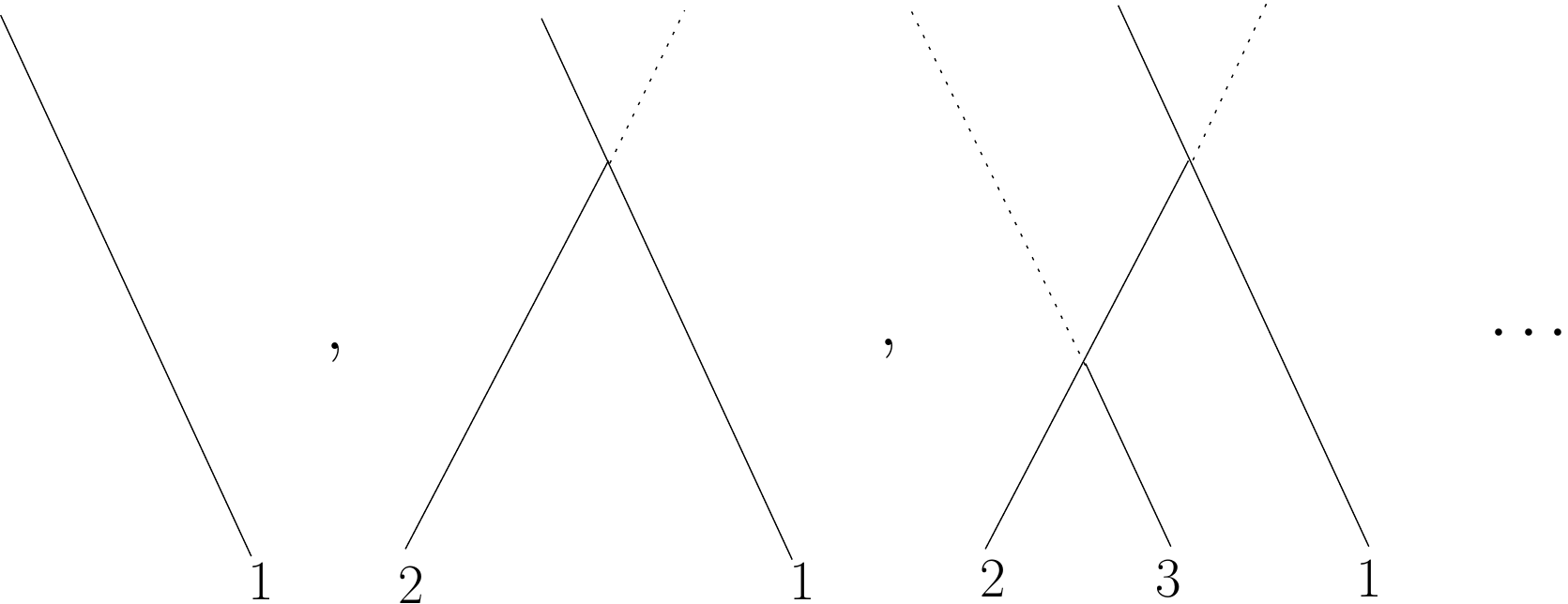} \label{eq:figI2b}
\ee
where the dotted lines correspond to free--flight. However, this example
is not enough, since we need to take into account additional trajectories, e.g.
\be
\includegraphics[width=4in]{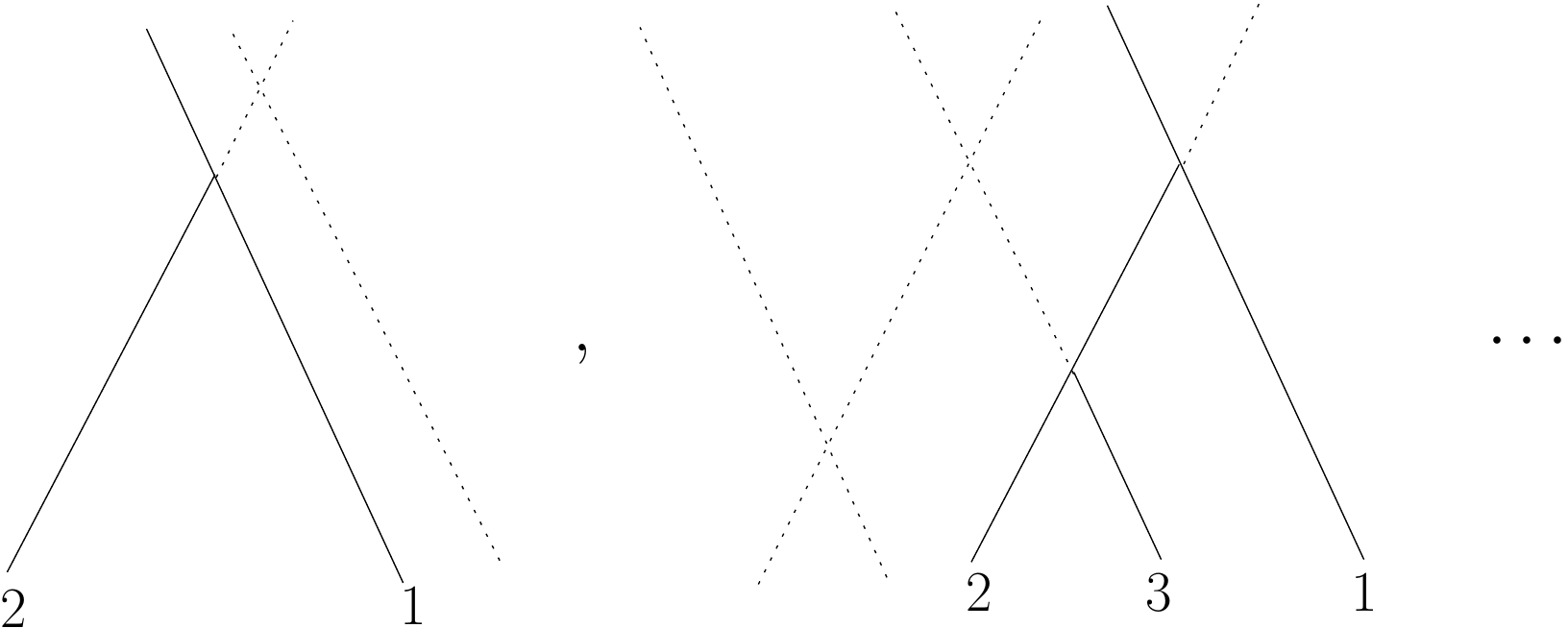} \label{eq:figI2c}
\ee

In other words, we {\em extend} the history of the backward cluster to provide
full knowledge of the trajectory of the particles in the time interval $(0,t)$. 

We make this more precise in the rest of this Section. Our goal is to write a formula for
the density of Bogolyubov clusters (Definition \ref{def:DBC} below) starting from \eqref{eq:fjexp2}
\& \eqref{eq:pdtr}.
Before that, we need to introduce notions of `collision graph' and of the associated `trajectory of clusters'.
The density of clusters will be indeed expressed as an integral over such trajectories.

Let $\GG_k$ be a labelled tree with
$k$ vertices, i.e. a connected graph with $k$ vertices and $k-1$ edges. For instance for $k=4$
\begin{center}
\includegraphics[scale=0.5]{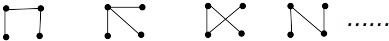}
\end{center}
The vertices are labelled $1,2,\cdots, k$ and $\GG_k = \{(i_1,j_1),\cdots,(i_{k-1},j_{k-1})\}$
(non ordered set of pairs).
Each vertex represents a particle and each link represents a collision. We refer to $\GG_k$
as {\em collision graph} of the Bogolyubov cluster.

A {\em trajectory of the Bogolyubov $t-$cluster} $(\bz_{k}(s))_{s\in (0,t)} = (z_1(s),\cdots,z_k(s))_{s\in (0,t)}$,
where $z_i(s) = (x_i(s),v_i(s)) = $ (position, velocity), is constructed as follows.

(a$'$) Fix $x_1,t,v_1,\cdots,v_k,\GG_k, t_1,\o_1, \cdots, t_{k-1},\o_{k-1}$, with $t_i \in (0,t)$;

(b$'$) let $\ell_1,\cdots,\ell_{k-1}$ be the permutation of $1,\cdots,k-1$ such that
 $t_{\ell_1} > t_{\ell_2} > \cdots > t_{\ell_{k-1}}$; 
construct the sequence of velocities $\bv^r$, $r=0, \cdots, k-1$, defined iteratively by:
$$
\bv^0=\bv_{k} = (v_1,\cdots,v_{k})\;,    \quad \bv^r=( v_1^{r-1},\cdots, v_{i_{\ell_r}}', \cdots,v_{j_{\ell_r}}',\cdots,v_{k}^{r-1})   \quad r\geq 1
$$
where, at step $r$,  the pair  $v_{i_{\ell_r}}'  , v_{j_{\ell_r}}'$ are the pre--collisional velocities 
(in the collision with impact vector $\o_{\ell_r}$) of the pair $v_{i_{\ell_r}}^{r-1},  v_{j_{\ell_r}}^{r-1}$ 
(assumed post--collisional), according to the transformation \eqref{eq:coll};

(c$'$) define the trajectory of particle $i\in \{1,2,\cdots,k\}$ of the Bogolyubov cluster iteratively by
$$
x_i(s) = x_i - v_i(t-s)\;, \quad v_i(s) = v_i \quad\quad\quad\quad\quad\quad s\in (t_{\ell_1},t)\;,    $$ and, for $r\geq 1$,
$$ x_i(s) = x_i(t_{\ell_r})-v_i^r(t_{\ell_r}-s)\;, \quad v_i(s) = v_i^r \quad\quad\quad\quad\quad\quad s\in (t_{\ell_{r+1}},t_{\ell_r})\;,
$$
with the convention $t_{\ell_k} = 0$. Notice that the positions $x_2,\cdots,x_k$ at time $t$ are uniquely
determined as soon as we fix $x_1(t) = x_1$.

With respect to the definition of trajectory used in \eqref{eq:fjexp2}, the essential differences
are summarized in the following table.
\begin{table}[htb]
\centering
\begin{tabular}{  l | c | r | rr}
  & sequence of collisions & times of collisions \ \ \ \ \ & \multicolumn{1}{r}{history of particle $i$} \\ \hline
  Backward cluster, size $k$ & tree graph $\G_{k-1}$ &  $t  > t_1 > \cdots > t_{k-1} >  0$\ \ &
  \multicolumn{1}{r}{\ \ specified in $(0,t_{i-1})$} \\ \hline
  Bogolyubov cluster, size $k$ \ \ & tree graph $\GG_k$ & $(t_1,\cdots,t_{k-1}) \in (0,t)^{k-1}$
  &\multicolumn{1}{r}{specified in $(0,t)$}
  \end{tabular}
\end{table}

\noindent In both cases, there are exactly $k-1$ collisions, since recollisions are forbidden
in the assumed kinetic regime (see page \pageref{point:rec}, item (2) for a precise statement). 
Such collisions are specified, respectively, by the {\em binary} tree graph
$\G_{k-1}$ and by the ordinary tree graph $\GG_k$.

The Bogolyubov cluster can be described as the time--symmetrized version of the backward cluster.
Alternatively, this can be seen as a symmetrization in the labelling of the particles, since no particle
plays a special role anymore.

\subsection{Size distribution of kinetic clusters} \label{sec:dc4}

Motivated by the previous discussion, we introduce here an explicit $f_t(k)$ written in terms of the Boltzmann
density solving \eqref{BE}, which should be interpreted physically as the fraction of particles of the gas belonging 
to a dynamical cluster of size $k$.  A formal argument {\em identifying} $f_t(k)$ as the kinetic limit of the corresponding quantity in a system of finitely many 
particles will be presented below (see Claim \ref{cl:FBC}); a rigorous derivation remains an open problem.

\begin{defi}\ \label{def:DBC}
Let $t\in[0,\infty)$ and $k\in\NNN$, then
\bea
&& f_t(k) := \frac{1}{(k-1)!}\sum_{\GG_k}  \int_{\L}dx_1 \int_{\RRR^{kd}} d\bv_{k} 
\int_{(0,t)^{k-1}} d\bt_{k-1} \, 
\prod_{r=1}^{k-1}\int_{S^{d-1}} d\o_{\ell_r} \,B(v^{r-1}_{i_{\ell_r}} -v^{r-1}_{j_{\ell_r}}\,,\,\o_{\ell_r})\,
\nn\\
&& \ \ \ \ \ \ \ \ \ \ \ \ \ \ \ \ \ \ \ \ \ \ \ \cdot \prod_{r=0}^{k-1}\,e^{-\int_{t_{\ell_{r+1}}}^{t_{\ell_r}} ds\, R_{k}(\bz_k(s),s)}
f_{0}^{\otimes k} (\bz^{k}(0))\;, \label{eq:ftk}
\eea
with the conventions $t_{\ell_0} = t, t_{\ell_k} = 0$.
\end{defi}

Compared to $f^{\G_{k-1}}$, see \eqref{eq:fjexp2} and \eqref{eq:pdtr}, $\int_0^t dt_1\int_0^{t_1} dt_2  \cdots \int_0^{t_{k-2}}dt_{k-1}$
has been replaced by the symmetric integral $\frac{1}{(k-1)!}\int_{(0,t)^{k-1}} d\bt_{k-1}$, and the 
integrand function is now the probability density of a trajectory of a Bogolyubov cluster with collision graph $\GG_k$.

By looking at \eqref{eq:ftk} in a simple example, we will show in Section \ref{sec:Max}
that this distribution is not normalized for all times. This is due to the development of giant clusters ($k = \infty$)
at some critical time $t_c$. Therefore, with respect to backward clusters
(often associated to the description of correlations \cite{APST14}), the Bogolyubov clusters exhibit a 
more interesting statistics.

The following rescaled version is also relevant.
\begin{defi}\ \label{def:FBC}
The kinetic fraction of Bogolyubov $t-$clusters with size $k$ is
\bea
&& g_t(k) := \frac{1}{Z_t}\,\frac{1}{k!}\,\sum_{\GG_k}  \int_{\L}dx_1 \int_{\RRR^{kd}} d\bv_{k} 
\int_{(0,t)^{k-1}} d\bt_{k-1} \, 
\prod_{r=1}^{k-1}\int_{S^{d-1}} d\o_{\ell_r} \,B(v^{r-1}_{i_{\ell_r}} -v^{r-1}_{j_{\ell_r}}\,,\,\o_{\ell_r})\,
\nn\\
&& \ \ \ \ \ \ \ \ \ \ \ \ \ \ \ \ \ \ \ \ \ \ \ \cdot \prod_{r=0}^{k-1}\,e^{-\int_{t_{\ell_{r+1}}}^{t_{\ell_r}} ds\, R_{k}(\bz_k(s),s)}
f_{0}^{\otimes k} (\bz^{k}(0))\;,\label{eq:gtk}
\eea
where 
\bea
&& Z_t := \sum_{k \geq 1} \frac{1}{k!}\,\sum_{\GG_k}  \int_{\L}dx_1 \int_{\RRR^{kd}} d\bv_{k} 
\int_{(0,t)^{k-1}} d\bt_{k-1} \, 
\prod_{r=1}^{k-1}\int_{S^{d-1}} d\o_{\ell_r} \,B(v^{r-1}_{i_{\ell_r}} -v^{r-1}_{j_{\ell_r}}\,,\,\o_{\ell_r})\,
\nn\\
&& \ \ \ \ \ \ \ \ \ \ \ \ \ \ \ \ \ \ \ \ \ \ \ \cdot \prod_{r=0}^{k-1}\,e^{-\int_{t_{\ell_{r+1}}}^{t_{\ell_r}} ds\, R_{k}(\bz_k(s),s)}
f_{0}^{\otimes k} (\bz^{k}(0))\label{eq:Zt}
\eea
is the normalization constant.
\end{defi}

The functions just introduced are the kinetic counterparts of the functions $f_t$ and $g_t$ examined
in \cite{GKBSZ08}. This connection with the finite dynamical system is clarified in the 
next Section.

%

\section{Derivation of (\ref{eq:ftk})} \label{sec:BG}


The following argument is an heuristic derivation of the formulas given above for the size distribution
of clusters. This is inspired by the papers \cite{Spohn,Si13}.

We consider here, for simplicity, a system of $N$ identical hard spheres.
These particles have unit mass and diameter $\e$ and move inside
the box $\L = [0,1]^3$ with reflecting boundary conditions.
The dynamics $\TT_N$ is given by free flow plus collisions at distance $\e$, 
which are governed by the laws of elastic reflection. 
We label the particles with an index $i=1,2,\cdots,N$.
The complete configuration of the system is then given by a vector $\bz_N=( z_1, \cdots, z_N )$, where 
$z_i=(x_i,v_i)$ collects position $x_i$ and velocity $v_i$ of particle $i$. 
Let us assign a probability density $W_0^N$ on the $N-$particle 
phase space, assuming it symmetric in the exchange of the particles.
For $j=1,2,\cdots,N$, we call $f^N_{0,j} = \int dz_{j+1} \cdots dz_N\, W_0^N(\bz_N)$ the $j-$particle marginal of $ W_0^N$.

Let $\TT_k, \TT'_{N-k}$ be the dynamical flows, considered in isolation, of the groups of particles $\{1,\cdots,k\}$ and $\{k+1,\cdots,N\}$
respectively. 
Set $\tilde\TT_N = (\TT_k, \TT'_{N-k})$ ($\tilde\TT_N = \TT_N$ only up to the time of the first collision
between the two groups).
We denote $s_i\in [0,\infty]$, $i=1,2,\cdots$ the time of the {\em first} collision of particle $k+i$ with the
set of particles $\{1,\cdots,k\}$ in the dynamics $\tilde\TT_N$.
Furthermore, we set $\tau_m = \min_{i\geq m} s_i$, that is the time of the first
collision of the group $\{k+m,\cdots,N\}$ with the group $\{1,\cdots,k\}$. 

Let us focus on Definition \ref{def:DBC} and 
let
\be
n_t(k) = \mbox{number of Bogolyubov $t-$clusters of size $k$}\;.
\ee
Moreover, let 
$$A := \Big\{``\{1,\cdots,k\} \mbox{\ forms a $t-$cluster under the reduced flow $\TT_k$''}\Big\}\;.$$
By the symmetry in the particle labels, the average of $n_t(k)$ with respect to $W^N_0$ is
\bea
&& \langle n_t(k) \rangle = \binom{N}{k}\, P\Big(\{1,2,\cdots,k\} \mbox{\ is a $t-$cluster }\Big)\nn\\
&& \ \ \ \ \ \ \ \ \ \ \ \ \ =  \binom{N}{k}\, \int d\bz_N \,W^N_0(\bz_N) \,\chi(A)\,  \chi\left(\tau_1 > t\right)\nn\\
&& \ \ \ \ \ \ \ \ \ \ \ \ \ =  \binom{N}{k}\, \int d\bz_k \,\chi(A)\, \int dz_{k+1}\cdots dz_N \,W^N_0(\bz_N)\,\chi\left(\tau_1 > t\right)\nn\\
&& \ \ \ \ \ \ \ \ \ \ \ \ \ \equiv  \binom{N}{k}\, \int d\bz_k \,\chi(A)\, f^N_{0,k}(\bz_k)\, P\left(\tau_1 > t \mid \bz_k\right) \;,
\label{eq:avntkst}
\eea
where $\chi(A)$ is the characteristic function of the set $A$, and the last line in \eqref{eq:avntkst} defines
the conditional probability.

More generally, we indicate by $P\left(\cdot \mid t_i,\cdots,t_1, \bz_k\right)$ a conditional probability
given $\bz_k$ and $s_{i} = t_{i},\cdots, s_1 = t_1$. Similarly,
$p_{s_i}(\cdot | t_{i-1},\cdots, t_1,\bz_k)$ is the conditional probability density of $s_i$, 
given $\bz_k$ and $s_{i-1} = t_{i-1},\cdots, s_1 = t_1$.
Using again the symmetry, 
\bea
&& P\left(\tau_1 > t \mid \bz_k\right)=  1-  P\left(\tau_1 < t \mid \bz_k\right) \nn\\
&& = 1- (N-k)\, \int_0^t dt_1\, p_{s_1}\left( t_1 \mid \bz_k\right)
\, P(  \tau_2 > t_1 \mid t_1,\bz_k) \nn\\
&& = 1- (N-k)\, \int_0^t dt_1\, p_{s_1}\left( t_1 \mid \bz_k\right)\nn\\
&& \ \ \ \cdot \Big( 1-
(N-k-1)\,\int_0^{t_1} dt_2\, p_{s_2}\left( t_2 \mid t_1, \bz_k\right)
 \, P(\tau_3>t_2 \mid t_2, t_1, \bz_k)\Big)\nn\\
&& = \cdots\;,
\eea
so that by iteration we obtain
\bea
&& P\left(\tau_1 > t \mid \bz_k\right) = \sum_{j=0}^{N-k} (-1)^j\,
(N-k)\,\int_0^t dt_1\, p_{s_1}(t_1 | \bz_k)\,\nn\\
&&\ \ \ \ \cdot (N-k-1)\,\int_0^{t_1} dt_2 \, p_{s_2}(t_2 | t_1,\bz_k) 
\cdots (N-k-j+1)\,\int_0^{t_{j-1}}dt_j
\,p_{s_j}(t_j | t_{j-1},\cdots, t_1,\bz_k)\;.\nn\\
\label{eq:Pt1cK}
\eea
The term with $j=0$ is defined to be $1$.

Next we evaluate $p_{s_i}(t_i | t_{i-1},\cdots, t_1,\bz_k)$.
Let $v_*$ be the velocity of particle $i$ at time $s_i$, $k_i \in \{1,\cdots,k\}$ the index of the particle that particle $i$
will hit, and $\o\in S^2$ the normalized relative displacement of particle $i$ with respect to
particle $k_i$. Let $z^{(k)}_{k_i}(s) = (x^{(k)}_{k_i}(s),v^{(k)}_{k_i}(s))$ be position and velocity of
particle $k_i$ along the flow $\TT_k$. By definition of $s_i$, we must find particle $i$ in a cylinder
of volume $ds_i\, dv_* d\,\o\, \e^2\, B(v^{(k)}_{k_i}(s_i)-v_*,\o)$ for some $k_i \in \{1,\cdots,k\}$. 
We remind that, for hard spheres, the cross--section is $B(V,\o) = \left(V \cdot \o\right)^+$.
It follows that, {\em assuming} the Boltzmann approximation $f^N_1 \sim f$ and $p_{s_i}(t_i | t_{i-1},\cdots, t_1,\bz_k) \sim 
p_{s_i}(t_i | \bz_k)$\footnote{As follows from the statistical independence of particle $i$
from {\em any} finite collection of given particles, in the limit $N \to \infty$.}, the conditional 
probability $dt_i\,p_{s_i}(t_i | t_{i-1},\cdots, t_1,\bz_k)$ will be close to $$\e^2 \sum_{k_i=1}^k  dt_i \int dv_* d\o \, 
B(v^{(k)}_{k_i}(t_i)-v_*,\o) \, f(x^{(k)}_{k_i}(t_i),v_*,t_i)\;.$$

Hence, in the scaling \eqref{eq:BG} we expect
\bea
&& P\left(\tau_1 > t \mid \bz_k\right) \approx \sum_{j \geq 0} (-1)^j
\prod_{i=1}^j \sum_{k_i=1}^k \,\int_{0}^{t_{i-1}} dt_i \int dv_* d\o \, 
B(v^{(k)}_{k_i}(t_i)-v_*,\o) \, f(x^{(k)}_{k_i}(t_i),v_*,t_i) \nn\\
&& \ \ \ \ \ \ \ \ \ \ \ \ \ \ \ \ \ \ \ = \sum_{j\geq 0}\frac{(-1)^j}{j!} \,\left(\int_0^t ds
\,\sum_{i=1}^k\,R(x^{(k)}_{i}(s),v^{(k)}_{i}(s),s)\right)^j\nn\\
&& \ \ \ \ \ \ \ \ \ \ \ \ \ \ \ \ \ \ \ = e^{-\int_0^t ds \,R_k(\bz^{(k)}_k(s),s)}
\eea
where $t_0 = t $ and we used the notations \eqref{nu}-\eqref{eq:Rjdef}.

Inserting this into \eqref{eq:avntkst}, we find
\be
\langle n_t(k) \rangle \approx \binom{N}{k}\, \int d\bz_k \,\chi(A)\, f^N_{0,k}(\bz_k)\, e^{-\int_0^t ds \,R_k(\bz^{(k)}_k(s),s)}\;.
\ee
We further assume:

(1) the marginals of the initial distribution factorize in the Boltzmann--Grad limit:
$f_{0,k}^{N}(\bz_k) \to f_0^{\otimes k}(\bz_k)$;

\label{point:rec}
(2) we may neglect recollisions; namely, the lower order term in \eqref{eq:avntkst} is given by
trajectories of the flow $\TT_k$ showing {\em exactly} $k-1$ collisions.

\noindent If this is true, then $\chi(A) = \sum_{\GG_k}\chi(\GG_k)$ where the tree graph $\GG_k = \{(i_1,j_1),\cdots,(i_{k-1},j_{k-1})\}$
specifies which pairs of particles collide in the $t-$cluster. It follows that
\be
\langle n_t(k) \rangle \approx \binom{N}{k}\, \sum_{\GG_k}\,
\int d\bz_k \,\chi(\GG_k)\, e^{-\int_0^t ds \,R_k(\bz^{(k)}_k(s),s)}\,f_0^{\otimes k}(\bz_k)\;.
\ee

Applying $k-1$ collision transforms, the measure can be rewritten as
\be
d\bz_k\,\chi(\GG_k) = \e^{2(k-1)} dx_1(t) dv_1(t)dv_2(t)\cdots dv_k(t)\, \prod_{r=1}^{k-1}dt_r d\o_r \,B(V_r\,,\,\o_r)\;,
\ee
where $x_1(t), v_1(t),\cdots,v_k(t)$ are position of particle $1$ and velocities of the particles 
of the cluster at time $t$,
while $t_1,\cdots,t_{k-1}$ are the collision times, $\o_r$ is the relative distance of $j_r$ and $i_r$ at the collision
time $t_r$, $V_r$ is the outgoing relative velocity at collision, and $B(V_r\,,\,\o_r)$ the corresponding cross--section
(see \cite{Si13} for a similar computation).

In the notations of the previous Section, this implies
\bea
&& \langle n_t(k) \rangle \approx  \binom{N}{k}\, \e^{2(k-1)}\,\sum_{\GG_k}\, \int_{\L}dx_1 \int_{\RRR^{3k}} d\bv_{k} 
\int_{(0,t)^{k-1}} d\bt_{k-1} \, \nn\\
&& \ \ \ \ \ \ \ \ \ \ \ \ \ \ \ \ \ \ \cdot 
\prod_{r=1}^{k-1}\int_{S^{d-1}} d\o_{\ell_r} \,B(v^{r-1}_{i_{\ell_r}} -v^{r-1}_{j_{\ell_r}}\,,\,\o_{\ell_r})\,
\prod_{r=0}^{k-1}\,e^{-\int_{t_{\ell_{r+1}}}^{t_{\ell_r}} ds\, R_{k}(\bz_k(s),s)}
f_{0}^{\otimes k} (\bz^{k}(0))\;.\nn\\
\eea
By \eqref{eq:BG}, $\binom{N}{k} \simeq \e^{-2k}/k!$. Therefore, we conclude that
\bea
&& \langle n_t(k) \rangle \approx  \frac{\e^{-2}}{k!}\,\sum_{\GG_k}\, \int_{\L}dx_1 \int_{\RRR^{3k}} d\bv_{k} 
\int_{(0,t)^{k-1}} d\bt_{k-1} \, \nn\\
&& \ \ \ \ \ \ \ \ \ \ \ \ \ \ \ \ \ \ \cdot 
\prod_{r=1}^{k-1}\int_{S^{d-1}} d\o_{\ell_r} \,B(v^{r-1}_{i_{\ell_r}} -v^{r-1}_{j_{\ell_r}}\,,\,\o_{\ell_r})\,
\prod_{r=0}^{k-1}\,e^{-\int_{t_{\ell_{r+1}}}^{t_{\ell_r}} ds\, R_{k}(\bz_k(s),s)}
f_{0}^{\otimes k} (\bz^{k}(0))\;.\nn\\
\eea

This heuristic discussion leads to the following claim.
\begin{cla}\ \label{cl:FBC}
Assume $f^N_1(t) \to f(t)$ as $\e\to 0$, where $f$ is a solution of
the Boltzmann equation with initial datum $f_0$. The
average density of clusters with size $k$ at time $t$, $f^N_t(k) := \frac{k \langle n_t(k) \rangle }{N}$
satisifies
\be
f^N_t(k) \longrightarrow f_t(k)
\ee
as $\e\to 0$, where $f_t(k)$ is given by \eqref{eq:ftk}.
\end{cla}
\noindent This is a law of large numbers type of result. 

Similarly, 
the total number of clusters $N_c(t):=\sum_{k=1}^{N}n_t(k)$
has fractional average 
\be
\frac{\langle N_c(t) \rangle }{N} : =\frac{1}{N} \langle \sum_{k=1}^{N}n_t(k)\rangle  \longrightarrow Z_t
\ee
given by \eqref{eq:Zt}, while the average fraction of clusters 
\be
g^N_t(k) := \langle \frac{n_t(k)}{N_c(t)} \rangle  \longrightarrow g_t(k)
\ee
given by \eqref{eq:gtk}.

Note that 
\be
f^N_t(k) = \frac{k\, g^N_t(k)}{\sum_{k \geq 1} k\, g^N_t(k)}\;.
\ee
This relation, however, breaks in the Boltzmann--Grad limit. In fact, we will show in the
following section that $f_t(k)$ is not normalized.

It would be interesting to write a rigorous derivation of \eqref{eq:ftk} along the above 
or similar arguments, using the known results of convergence to the Boltzmann equation.
This will be matter of further investigation.
An exact perturbative expression for the path measure of one tagged sphere in a gas at thermal
equilibrium can be found in \cite{LS82}.

\section{Phase transition for Maxwell molecules} \label{sec:Max}

It has been observed in Reference \cite{GKBSZ08} that the $t-$cluster size distribution of 
an elastic billiard consisting of $N=10^4$ disks with volume density $10^{-3}$ is well described
by a power law with exponential damping
\be
k^{-5/2}e^{-k/\g(t)} \;.
\label{eq:LawS}
\ee
The strength of damping $\g(t)^{-1}$ turns out to be monotone decreasing for small times and such that
$\g(t)^{-1} \to~0 $ as $t \to t_c$. Close to the critical time $t_c$, the distribution transforms sharply
from exponential to pure power law. Moreover, after the critical time, a distinct gap appears between
the largest cluster and the rest of the distribution. The maximal cluster starts to dominate as in a percolation
model.

Consider now the Boltzmann equation model
\be
\pa_t f= J(f,f)-f
\label{BEM}
\ee
where $f = f(v,t)$,
\be
J(f,f)(v)=
\int_{\RRR^d\times S^{d-1}} dv_1 d\o \ g\left(\cos\theta\right) f(v_1')f(v')
\ee
for some nonnegative function $g$ satisfying $g=0$ for $\cos\theta < 0$, 
and $$\cos\theta = \frac{(v-v_1)}{|v-v_1|}\cdot\o\;.$$
We have fixed the time scale in such a way that 
\be
\int_{ S^{d-1}}  d \o \,   g(\cos\theta)=1\;.
\label{norm}
\ee

The distribution of clusters is obtained from the formulas in Section \ref{sec:dc4},
by inserting the kernel $B(v-v_1,\o) = g(\cos\theta)$ and 
noticing that, in this case, the normalization of $f$ implies $R_k = k$ (see \eqref{eq:Rjdef}--\eqref{nu}).
Definition \ref{def:FBC} leads to
\be
g_t(k) = \frac{1}{Z_t}\frac{k^{k-2}}{k!} t^{k-1} e^{-kt}\;,
\label{eq:gtkMaxe}
\ee
having used that the number of labelled trees with $k$ vertices is
$\sum_{\GG_k}1=k^{k-2}\;.$ Note that no dependence on the 
dimension or on the initial data is left.

Stirling's approximation gives
\be
g_t(k) \simeq \frac{1}{Z_t}\, \frac{(ete^{-t})^{k}}{\sqrt{2\p} \, t \, k^{5/2}}\;.
\ee
The function $ete^{-t}$ is strictly smaller than $1$ for $t \neq 1$.
The function 
\be
Z_t = \sum_{k \geq 1}\frac{k^{k-2}}{k!} t^{k-1} e^{-kt}
\ee
is monotonic decreasing with $Z_0 = 1$ and $Z_t \simeq e^{-t}$ for $t$ large. 
Hence $g_t(k) \sim e^{-(k-1)t}$ for $t$ large.

The behaviour of $Z_t$ and $g_t(k)$ is well known.
For $t\in [0,1]$, the function $Z_t\, g_t(k)$ is the solution of the Smoluchowski coagulation
equation with product kernel, and at $t=1$ a critical transition occurs known as {\em gelation};
see e.g. \cite{Fl41,St43,McL62,EZH83,Al99,Re13}. 

Summarizing, the following picture emerges.

\begin{itemize}
\item At time zero $g_t(k) = \d_{k,1}$. All the clusters have size $1$.
\item For $0\leq t \leq 1$, the distribution changes smoothly. 
As the model evolves, clusters of size $k\geq 2$ are formed, while the fractional total number of clusters 
decreases. The fraction of singletons $g_t(1)$ also decreases. At fixed time, $g_t(k)$ is a power law with 
exponential damping. This reproduces the law \eqref{eq:LawS} measured in \cite{GKBSZ08},
with $\g(t)= |\ln(ete^{-t})|^{-1}\;.$
\item At the critical time $t_c = 1$, $g_t(k)$ becomes a pure power law with exponent $5/2$. This is a phase
transition of the second kind. The second derivative of $Z_t$ is zero for $t<t_c$ and
has a positive jump at $t_c$.
\item For $t>t_c$, the distribution changes again smoothly, but with inverse behaviour. 
The fraction of clusters of size $k\geq 2$ decreases, while the fraction of singletons increases. 
The law \eqref{eq:LawS} is still verified with same $\g(t)$.
\end{itemize}

The phase transition described above is associated to the appearance of a
{\em giant cluster} with $k = \infty$. 

The presence of such a giant component is directly visible by looking at the mass density of clusters,
Definition \ref{def:DBC}, which in the model of this Section becomes
\be
f_t(k) = \frac{k^{k-2}}{(k-1)!} t^{k-1} e^{-kt}\;.  \label{eq:ftkMaxexa}
\ee
Observe that $f_t(1) = e^{-t}$, $f_0(k) = \d_{k,1}$ and 
\be
f_t(k) \simeq \frac{(ete^{-t})^{k}}{\sqrt{2\p} \, t \, k^{3/2}}
\ee
for $k$ large enough.

The total mass of clusters is defined as the normalization constant of $f_t(k)$:
\be
F_t = \sum_{k\geq 1} f_t(k)= \sum_{k\geq 1}\frac{k^{k-2}}{(k-1)!} t^{k-1} e^{-kt}\;.
\ee
There holds $F_t = 1$ for $t\leq t_c=1$,  $F_t < 1$ for $t > 1$ and $F_t \sim e^{-t}$ for $t$ large.
This leads to define the ``density of giant clusters'' for the Boltzmann equation as
$ F^{\infty}_t = 1-F_t\;.$

\bigskip

\noindent 
{\bf Acknowledgments.} We thank M. Pulvirenti, Ya. Sinai and H. Spohn for discussions and suggestions.
We are particularly grateful to J. J. L. Vel\'{a}zquez for having pointed out the connection with coagulation
equations. 

S.S. has been supported by Indam--COFUND Marie Curie fellowship 2012, call 10.

Finally, we thank the anonymous referee for his valuable suggestions, that led to considerable
improvement of this paper.

\end{document}